\documentclass[12pt]{article}  % Changed from 11pt to 12pt for larger base font size
\usepackage[margin=1in]{geometry}

\usepackage{anyfontsize}
\usepackage{setspace}
\usepackage{graphicx}
\usepackage{newtxtext}
\usepackage{bm}
\usepackage{booktabs}
\usepackage{natbib}
\usepackage{amsmath, amssymb}
\usepackage{mathrsfs}
\usepackage{mathtools}
\usepackage{hyperref}
\hypersetup{
    colorlinks = true,
    urlcolor   = blue,
    citecolor  = black,
}

\usepackage{authblk}  % For author and affiliation formatting

\title{\textbf{Thermodynamically Admissible Diffuse Interface Model for Nanoscale Transport of Dense Fluids}}

\author[1]{Rahul Bhattacharjee}
\author[2]{Henning Struchtrup}
\author[1]{Anirudh Singh Rana\thanks{Corresponding author: \texttt{anirudh.rana@pilani.bits-pilani.ac.in}}}

\affil[1]{Department of Mathematics, Birla Institute of Technology and Science Pilani, Rajasthan 333031, India}
\affil[2]{Department of Mechanical Engineering, University of Victoria, PO Box 1700 STN CSC, Victoria, V8W 2Y2, British Columbia, Canada}

\date{}  % You can insert \today if you want the date shown

\renewenvironment{abstract}
  {\small\quotation\noindent\fontsize{12.5pt}{15pt}\selectfont\textbf{Abstract.} }
  {\endquotation}

\begin{document}

\maketitle

\begin{abstract}
We investigate interfacial fluid dynamics and heat transfer at nanoscales using an improved diffuse interface approach for liquid-vapor interfaces in non-equilibrium. Conventional Navier-Stokes-Korteweg (NSK) formulations often fail to accurately capture transport phenomena across extremely thin interfaces due to underestimation of interface resistances. In this work, we improve the NSK model by adding a production term in the momentum equation based on higher-order corrections. To enhance interface resistances, viscosity and thermal conductivity are made dependent on the density gradient, increasing resistance only within the interface region. The gradient-based coefficients are determined by fitting to solutions of the Enskog–Vlasov (EV) equation for Couette flow \cite{Struchtrup_Frezzotti_2022}. Applying these fitted equations to pure heat conduction and planar evaporation problems shows that the model accurately captures interfacial transport, making it a useful tool for studying nanoscale evaporation, thermal management, and droplet dynamics on solid surfaces.
\end{abstract}

% \begin{keywords}
% Enskog-Vlasov, Navier-Stokes-Korteweg, Couette flow, Vlasov force.
% \end{keywords}
% {\bf MSC Codes }  {\it(Optional)} Please enter your MSC Codes here
\section{Introduction}
\label{sec:headings}
Nanoscale phase transition plays a vital role in modern thermal management systems, which are essential in applications ranging from high-performance electronics and energy systems to advanced manufacturing \citep{gong2024mesoscopic}. Technologies such as heat pipes, microchannel evaporators, phase-change cooling, and thin-film boiling devices rely heavily on accurate modeling of phase transition at liquid-vapor interfaces \citep{dupont2004heat,mukherjee2009contribution,rana2019lifetime}. As device sizes shrink and localized heat flux increases, predicting interfacial transport processes with high physical fidelity has become increasingly critical. Accurate modeling of such phenomena requires capturing the interplay between capillarity, surface forces, and transport processes within the interfacial region—a challenge that spans from the microscale to the nanoscale \citep{anderson1998diffuse}.

\textcolor{black}{A popular theoretical framework for simulating such multiphase systems \textcolor{black}{is provided by the Navier-Stokes-Korteweg (NSK) equations,}  which originated from the foundational ideas of \citet{gibbs1878equilibrium} on the concept of a dividing surface, and were later extended through the thermodynamic formulation of diffuse interfaces by \citet{van1979thermodynamic} and \citet{korteweg1901forme}. These pioneering studies established that interfaces possess finite thickness, with van der Waals notably demonstrating how this thickness diverges near critical temperatures using his equation of state. Building upon these classical foundations, modern implementations of the NSK equations augment the classical Navier-Stokes system with a Korteweg stress to represent capillarity in a diffuse-interface setting \citep{anderson1998diffuse}. In this formulation, the interface is treated as a finite region across which fluid properties vary smoothly, rather than as a sharp boundary \citep{10.57262/ade/1356651336}. Recent advances in numerical methods have significantly enhanced NSK modeling capabilities: \citet{gomez2010isogeometric}-- employed isogeometric analysis, while \citet{liu2013functional}-- developed a semi-discrete Galerkin method using functional entropy variables. Other notable contributions include \citet{diehl2016numerical}-- numerical approximation method on the basis of the Local Discontinuous Galerkin method, \citet{martinez2020high}-- high-order finite volume method for phase-change simulations, and \citet{tian2016h}-- adaptive local discontinuous Galerkin method for non-isothermal flows. Contemporary work by \citet{souvcek2020thermodynamic}-- on second-law-compliant boundary conditions and \citet{dhaouadi2023structure}-- structure-preserving finite volume scheme further demonstrate the ongoing development of this framework. Despite these advances, the standard NSK formulation assumes density dependent transport coefficients and omits nonlocal corrections that become crucial when the interfacial thickness is small—a common situation in nanoscale evaporation and droplet spreading \citep{anderson1998diffuse}.}
Building on the theoretical framework described above, we propose two significant enhancements to the NSK model. First, we introduce \textcolor{black}{a correction term for intermolecular transfer in the momentum equation} from higher-order corrections to the momentum balance formulation developed by \citet{Struchtrup_Frezzotti_2022}. This modification enables more precise characterization of momentum exchange at interfaces. Second, we implement density-gradient-dependent transport coefficients---specifically thermal conductivity and viscosity---following the approaches established by \citet{bedeaux2003nonequilibrium,johannessen2004nonequilibrium}.  This implementation better accounts for the influence of microscopic fluid structure on transport phenomena within interfacial regions. Indeed, the sudden density change leads to interfacial resistances that are not captured in the classical Korteweg-type models. Thus, in this study, we develop a modified NSK formulation referred to as the Augmented Navier-Stokes-Korteweg model (ANSK) for the remainder of the paper, which specifically addresses the limitations of conventional models when applied to nanoscale interfaces, where interfacial thickness and non-local effects become particularly significant.

Further, the effectiveness of the ANSK model is assessed by studying three fundamental problems: Couette flow, pure heat conduction, and evaporation from a planar surface. Our results demonstrate \textcolor{black}{the gradient dependent coefficients effectively increase resistance only in the interface. The coefficients in this phenomenological approach are fitted by comparison with solutions of the EV equation-- \cite{Struchtrup_Frezzotti_2022}, for Couette flow. Application of the fitted equations to heat and mass transfer problems yields excellent agreement to EV simulations---a kinetic theory based approach known for its accuracy but high computational cost \citep{frezzotti1993numerical,wu2015fast}.} Thus, our ANSK model offers a balanced framework that combines thermodynamic consistency, computational efficiency, and improved fidelity for nanoscale transport processes in dense fluids. 

\textcolor{black}{The structure of the remainder of the paper is as follows: Section \ref{sec:level2} presents the conservation laws and their corresponding closure relations. In Section \ref{sec:level3}, the ANSK model is introduced. Section \ref{sec:level4} is dedicated to the presentation and analysis of three benchmark problems. Finally, concluding remarks are provided in Section \ref{sec:level5}.}

\section{\label{sec:level2}Conservation laws from the Enskog-Vlasov Equation}
We begin with the conservation laws for mass, momentum, and energy derived from the Enskog–Vlasov (EV) equation \citep{grmela1971kinetic,sobrino1967kinetic,Struchtrup_Frezzotti_2022}. Let $\rho$, $\bm{v}$, $\theta$, $\bm{\sigma}$ and $\bm{q}$ respectively denote mass density, velocity, temperature (in energy units), trace-free kinetic stress tensor and kinetic heat flux. The conservation equations read:
\begin{subequations}
\label{Conservation laws}
\begin{align}
\frac{D\rho}{Dt} + \rho \nabla \cdot \bm{v} = 0, \quad
\rho \frac{D\bm{v}}{Dt} + \nabla(\rho \theta) + \nabla \cdot \bm{\sigma} - \rho \bm{\mathcal{F}} = \bm{\Upsilon}^{1}, \\
\frac{3}{2} \rho \frac{D\theta}{Dt} + \rho \theta \nabla \cdot \bm{v} + \bm{\sigma} : \nabla \bm{v} + \nabla \cdot \bm{q} = \Upsilon^{2}.
\end{align}
\end{subequations}
The term 
$\bm{\mathcal{F}}$, represents the Vlasov force arising from long-range intermolecular interactions \citep{vlasov1961many}, while the transfer terms $\bm{\Upsilon}^{1}, \Upsilon^{2}$ account for the short-range collisional contributions due to finite-size particle effects and local spatial correlations, as described by the Enskog collision operator \citep{grmela1971kinetic}.

The Vlasov force can be expressed as the divergence of the Korteweg stress tensor $(\bm{\tau}^K)$ as $\rho \bm{\mathcal{F}} = -\nabla \cdot \bm{\tau}^K$,   
where $\bm{\tau}^K$ is given by 
\begin{align}
\bm{\tau}^K = 
- \frac{2\pi a^3}{3} \frac{\phi_a}{m} \chi_1 \rho^2 \, \mathbf{I}
- \frac{2\pi a^5}{15} \frac{\phi_a}{m} \chi_3 
\left[
\left( \rho \nabla^2 \rho + \frac{1}{2} |\nabla \rho|^2 \right) \mathbf{I}
- \nabla \rho \otimes \nabla \rho
\right].
\end{align}
Here $\mathbf{I}$ is the identity tensor, $\phi_a>0$, having the same dimensions as that of temperature appearing in Sutherland potential \citep{sutherland1893lii}, $\chi_{1}$, $\chi_{3}$ are the Korteweg coefficients and $a$, $m$ are the diameter and mass of the particle respectively. 
% Furthermore, the Korteweg coefficients $\chi_{1}$ and $\chi_{3}$ are defined as 
% \begin{align}
%     \chi_{1} = \frac{\gamma}{\gamma-3},~~~~~~~~~~{\chi}_{3}(\theta) = c_{0}+c_{1}\theta+c_{2}\theta^{2}+c_{3}\theta^{3},  
% \end{align}
% with \( \chi_1 = 2 \), corresponds to a power-law potential with \( \gamma = 6 \) and $\chi_{3}$, is expressed as a function of temperature for details interested readers refer \citep{bhattacharjee2024temperature}, further \[c_{0}=-2.25557, c_{1}=18.2905, c_{2}=-25.0087, c_{3}=18.1759.\]
\vspace{-0.4cm}
\subsection{Transfer terms from the Enskog collision operator}
To close the system in (\ref{Conservation laws}), constitutive relations and transfer terms must be expressed in terms of known variables. \textcolor{black}{Following the equilibrium formulation of \citet{wang2020kinetic}--- the momentum transfer term is modeled as
\begin{equation}
\bm{\Upsilon}^{1} = -\frac{2\pi a^3}{3m} \nabla (\rho^{2} \theta Y) 
\underbrace{-\frac{4\pi a^3}{15m} \nabla \cdot (\rho Y \bm{\sigma}) 
+ \nabla \cdot \bm{\Phi}^{(4)}},
\end{equation}
where the last two terms above underbraces have been modeled following \citet{Struchtrup_Frezzotti_2022}, which takes in to account the non-equilibrium effects.}
Here
\begin{align*}
   \bm{\Phi}^{(4)} = \frac{4 \sqrt{\pi} a^{4}}{15 m} \biggl[ 
\mathbf{I}  \rho^{2} Y \sqrt{\theta} (\nabla \cdot \mathbf{v})  
+  \rho^{2} Y \sqrt{\theta} \nabla \mathbf{v}  
+   \rho^{2} Y \sqrt{\theta} (\nabla \mathbf{v})^{\mathsf{T}}
\biggr],
\end{align*}
whose divergence is the second-order symmetric derivative of velocity---fourth order correction term in $a$, following \citet{Struchtrup_Frezzotti_2022}. The effect of this term considered in the ANSK model--shall be discussed in detail in sections to follow.
% In a similar manner the intermolecular energy  transfer term arising from the Enskog collision operator accounts for energy redistribution due to collisions and non-local effects following \citet{Struchtrup_Frezzotti_2022}, is modeled as
% \begin{equation}
% \Upsilon^{2} = - \frac{2\pi}{3} \frac{a^3}{m} \nabla \cdot \left\{ \rho Y \left[ \rho \theta \mathbf{v} + \frac{2}{5} \left( \boldsymbol{\sigma} \cdot \mathbf{v} \right) \right] \right\}.
% \end{equation}
\textcolor{black}{In a similar manner the energy transfer term following the equilibrium formulation of \citet{wang2020kinetic}, is modeled as
\vspace{-0.55cm}
\begin{align}
\Upsilon^{2} = -\frac{2 \pi}{3 }\frac{a^{3}}{m}\rho^{2} \theta Y(\nabla \cdot \mathbf{v}).
\end{align}}
Here, \( Y(\rho) \) is the pair correlation function at contact, given by \citet{carnahan1969equation}, equation as
\vspace{-0.3cm}
\[
Y = \frac{1}{2} \frac{2 - \eta}{(1 - \eta)^3}, \quad \text{with} \quad \eta = \frac{\pi a^3 \rho}{6}.
\]
\noindent\underline{\textbf{Second law of thermodynamics:}}~~Assuming entropy depends only on \( \rho \) and \( \theta \), we postulate the following functional form for the entropy
\begin{equation}
\varsigma(\rho, \theta) = \frac{3}{2} \ln \theta - \ln \rho - g(\rho), \quad
g(\rho) = \frac{2}{3} \frac{\pi a^3}{m} \int Y(\rho) \, d\rho.
\end{equation}
Using the Gibbs relation and substituting from the conservation laws, the entropy balance gives
\begin{equation}
\rho \frac{D\varsigma}{Dt} + \nabla \cdot \left( \frac{\bm{q}}{\theta} \right) = 
- \frac{1}{\theta} \bm{\sigma} : \nabla \bm{v} - \frac{1}{\theta^2} \bm{q} \cdot \nabla \theta = \Sigma.
\end{equation}

Entropy production is non-negative when the following constitutive relations hold:
\textcolor{black}{\begin{align}
\label{constitutive equations}
\boldsymbol{\sigma} = -\mu \left( \nabla \mathbf{v} + (\nabla \mathbf{v})^{\mathsf{T}} \right) + \frac{2}{3} \mu \left( \nabla \cdot \mathbf{v} \right) \mathbf{I}, \quad
\bm{q} = -\lambda \nabla \theta,
\end{align}}
where \( \mu>0 \), \( \lambda>0 \) are the shear viscosity and thermal conductivity respectively. 
Thus, entropy production \( \Sigma \geq 0 \) is guaranteed proving the second law of thermodynamics, and the entropy flux is identified as \( \bm{\Psi} = \bm{q}/\theta \). Together, the conservation laws (\ref{Conservation laws}), and constitutive relations (\ref{constitutive equations}) form a closed Navier--Stokes--Korteweg type system derived from the Enskog--Vlasov framework. 
\vspace{-0.6cm}
\section{\label{sec:level3}Augmented Navier-Stokes Korteweg Model}
The conservation laws for mass, momentum, and energy in conservative form are given by:
\begin{align}
\label{conform}
\frac{\partial \rho}{\partial t} + \nabla \cdot (\rho \bm{v}) &= 0, \quad
\rho \frac{D \bm{v}}{Dt} + \nabla \cdot \bm{\Pi} &= \bm{0},\quad
\rho \frac{D}{Dt} \left( \tfrac{3}{2} \theta + \tfrac{1}{2} |\bm{v}|^2 + \varepsilon_K \right) 
+ \nabla \cdot \bm{\mathcal{J}} &= 0,
\end{align}
where the total momentum and energy flux are defined as:
\begin{align}
\bm{\Pi} &= \left[ \rho \theta + \frac{2\pi a^3}{3m} \rho^2 \theta Y \right] \mathbf{I} 
+ \left(1 + \frac{4\pi a^3}{15m} \rho Y \right) \bm{\sigma} + \bm{\tau}^K -  \bm{\Phi}^{(4)}, \\
\label{enrflux}
\bm{\mathcal{J}} &= \bm{q} + \bm{q}^K + (\rho \theta \mathbf{I} + \bm{\sigma} + \bm{\tau}^K) \cdot \bm{v} 
+ \textcolor{black}{\underbrace{\frac{2\pi}{3}\frac{a^{3}}{m} \rho Y \left( \rho \theta \bm{v} + \frac{2}{5} \bm{\sigma} \cdot \bm{v} \right),}}
\end{align}
\textcolor{black}{where the term within the braces in \ref{enrflux} have been obtained following the formulation as modeled in \citet{Struchtrup_Frezzotti_2022}.}
The potential energy due to the attractive forces \citep{anderson1998diffuse} and the additional energy flux due to nonlocal effects \citep{giovangigli2020kinetic} are:
\begin{equation}
\varepsilon_K = -\frac{2\pi a^3}{3m} \chi_1 \phi_a \rho 
+ \frac{\pi a^5}{15m} \chi_3 \phi_a \frac{1}{\rho} |\nabla \rho|^2, \quad
\bm{q}^K = \frac{2\pi a^5}{15m} \chi_3 \phi_a \rho (\nabla \rho)(\nabla \cdot \bm{v}).
\end{equation}
\vspace{-0.08cm}
To incorporate interfacial and non-equilibrium effects, the transport coefficients—thermal conductivity \( \lambda \) and viscosity \( \mu \)—are modified to depend on both the local density and its gradient. Such dependencies become significant near phase interfaces, as discussed in \citep{bedeaux2003nonequilibrium,johannessen2004nonequilibrium}. The expressions for \( \lambda \) and \( \mu \) are given by:
\begin{align}
\label{Heat Conductivity new}
\lambda\left(\rho, |\nabla \rho|\right) &= \frac{15}{4} \frac{\rho \theta \left(1+\frac{2\pi}{5}\frac{a^{3}}{m}\rho Y\right)^{2}}{\frac{16}{5}\frac{\rho}{m}\sqrt{\pi \theta}a^{2}Y\underbrace{\left[1+ \alpha\left(\frac{a}{\rho}|\nabla \rho|\right)^{2}\right]^{2}}} 
+ \frac{2}{3} \frac{a^{4}\sqrt{\pi \theta}}{m}\rho^{2} Y, \\
\label{Coefficient of Viscosity new}
\mu\left(\rho, |\nabla \rho|\right) &= \frac{\rho \theta \left(1+\frac{4\pi}{15}\frac{a^{3}}{m}\rho Y\right)^{2}}{\frac{16}{5}\frac{\rho}{m}\sqrt{\pi \theta}a^{2}Y\underbrace{\left[1+ \alpha\left(\frac{a}{\rho}|\nabla \rho|\right)^{2}\right]^{2}}}
+ \frac{4}{15} \frac{a^{4}\sqrt{\pi \theta}}{m}\rho^{2} Y,
\end{align}
where the coefficient \( \alpha \)  is a temperature-dependent parameter, detailed in later sections. Interfacial modifications arising from density gradients are highlighted via the underbrace, marking non-equilibrium contributions in the interfacial region.

The ANSK model thus consists of the conservation laws \eqref{conform}, constitutive relations \eqref{constitutive equations}, and modified transport coefficients \eqref{Heat Conductivity new}–\eqref{Coefficient of Viscosity new}.
\vspace{-0.6cm}
\section{\label{sec:level4}Application of NSK Model on three benchmark problem}
\subsection{Two-Phase Isothermal Couette Flow}
To validate the ANSK model, we simulate steady, isothermal Couette flow across a liquid–vapour interface and compare results with DSMC benchmarks \citep{frezzotti2012slip,rah2001theory}. For the physical setup, we define a coordinate system where the interface normal is aligned with the \( x \)-axis. The liquid phase is located in the region \( \Omega = \{(x, y, z) \in \mathbb{R}^3 : 0 \leq x < x_l\} \), where \( x_l \) represents the thickness of the liquid film. The vapor phase occupies the region to the right of this interface. Flow is unidirectional parallel to the liquid-vapor interface with \( \boldsymbol{v}(x) = (0, v_y(x), 0) \), where \( v_y(0) = 0 \) and \( dv_y/dx \to \text{const} \) as \( x \to L \).
% \begin{figure}
%     \centering
%  \includegraphics[width=0.40\textwidth]{Schematic Couette flow.png}
%     \caption{Schematic of Isothermal Couette flow.}
% \end{figure}
The incompressible and steady flow satisfies \( \nabla \cdot \boldsymbol{v} = 0 \), so the governing variables reduce to \{\( \rho(x), v_y(x), \sigma_{xy}(x) \)\}. \textcolor{black}{For remainder of the article all variables are non-dimensionalized by setting \( a = m = \phi_a = 1 \). The ANSK model reduces to 
\begin{align}
\label{modifiedcouetteeqn}
\frac{d}{dx}\left\{
\left[1 + \frac{4\pi}{15}  \rho Y \right] {\sigma_{xy}}
- \frac{4 \sqrt{\pi}}{15} \rho^2 Y \sqrt{\theta} \frac{d v_{y}}{dx}
\right\} &= 0, \quad
\mu \frac{d v_{y}}{dx} + {\sigma_{{xy}}} = 0.
\end{align}}
The Korteweg coefficients are specified as $\chi_{1} =2$ and $\chi_{3}$ as a temperature dependent function defined as ${\chi}_{3}(\theta) = c_{0}+c_{1}\theta+c_{2}\theta^{2}+c_{3}\theta^{3}$, expressed as a function of $\theta$, for details refer \citep{bhattacharjee2024temperature}. Here $c_{0} = -2.25557$, $c_{1} = 18.2905$, $c_{2} = -25.0087$ and $c_{3} = 18.1759$.
Equations \eqref{modifiedcouetteeqn} are discretized using midpoint schemes \citet{ascher1998computer}, with a uniform mesh in \( x \in [0, L] \), with $L=50$.

\noindent\textbf{Boundary Conditions:} \hspace{1em}
\textcolor{black}{\( v_y(0) = 0, \quad
\left.\sigma_{xy} \right|_{x=L}= \text{assigned constant}, \quad
\left. \frac{d\sigma_{xy}}{dx} \right|_{x=0,L} = 0. \)}

Density and gradient-dependent coefficients use equilibrium values from \eqref{conform}, evaluated as in \citep{frezzotti2005mean,Struchtrup_Frezzotti_2022,bhattacharjee2024temperature}. The resulting system of linear equations is  solved numerically using MATLAB\textsuperscript{\textregistered}. The parameter \(\alpha\) is varied with the dimensionless temperature \(\theta\) by fitting it to match the EV equation solutions for the Couette flow problem, following \citet{Struchtrup_Frezzotti_2022}.
%\vspace{-0.9cm}
\subsubsection{Analysis of results of the two-phase Couette flow}
\begin{figure}
 \centering
 \includegraphics[width = 0.5\textwidth]{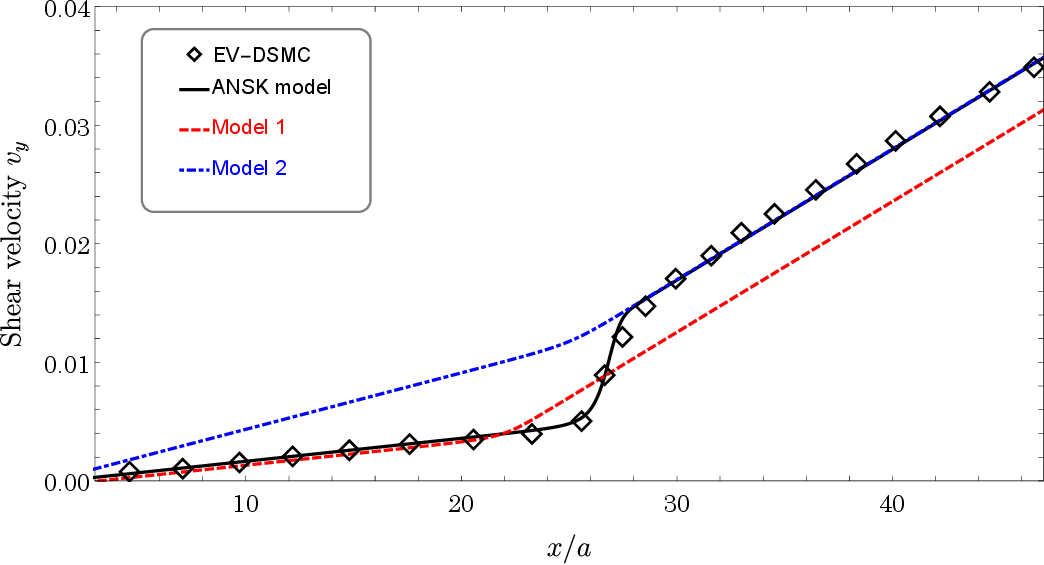}\hfill
 \includegraphics[width = 0.5\textwidth]{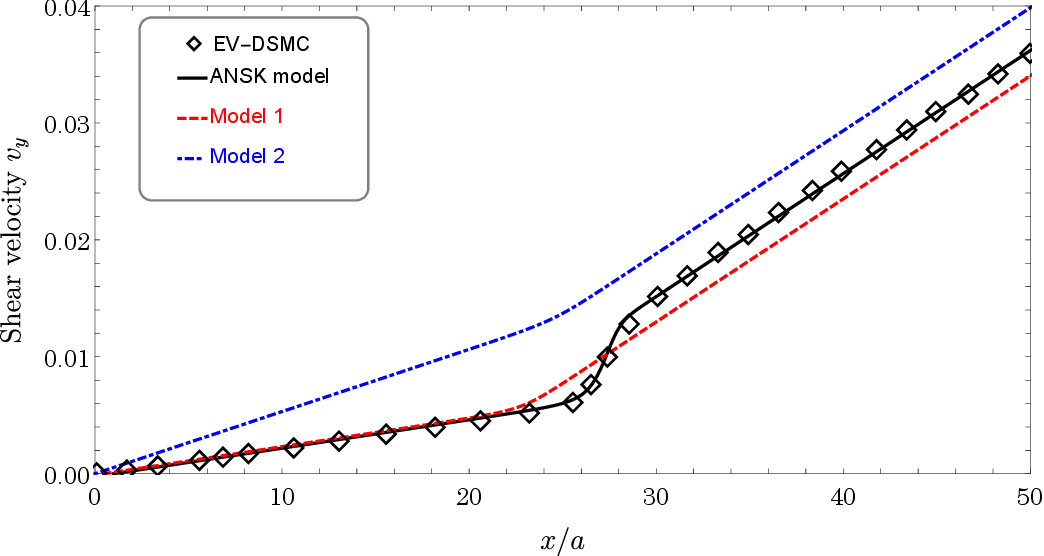}
 \includegraphics[width = 0.5\textwidth]{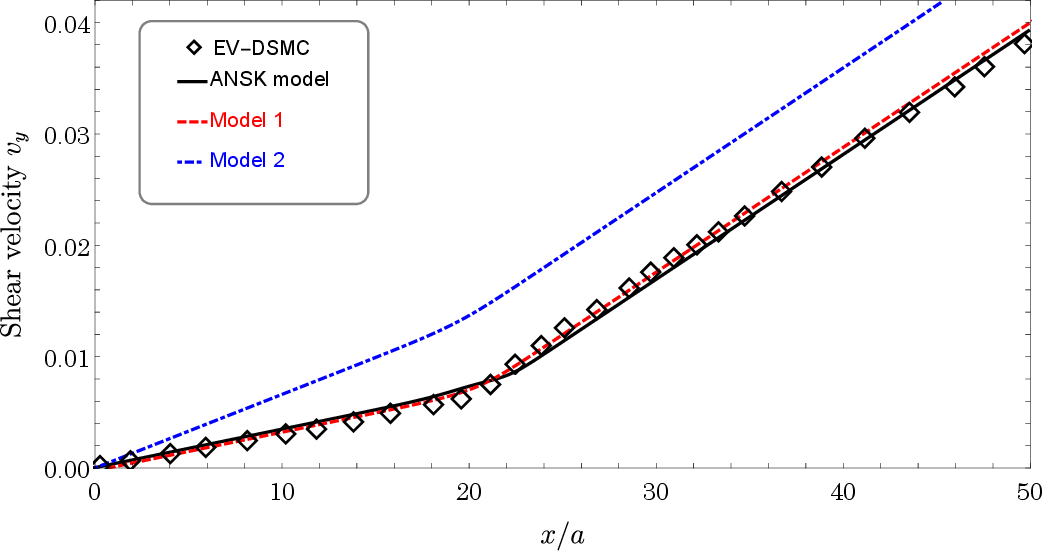}\hfill
 \includegraphics[width = 0.5\textwidth]{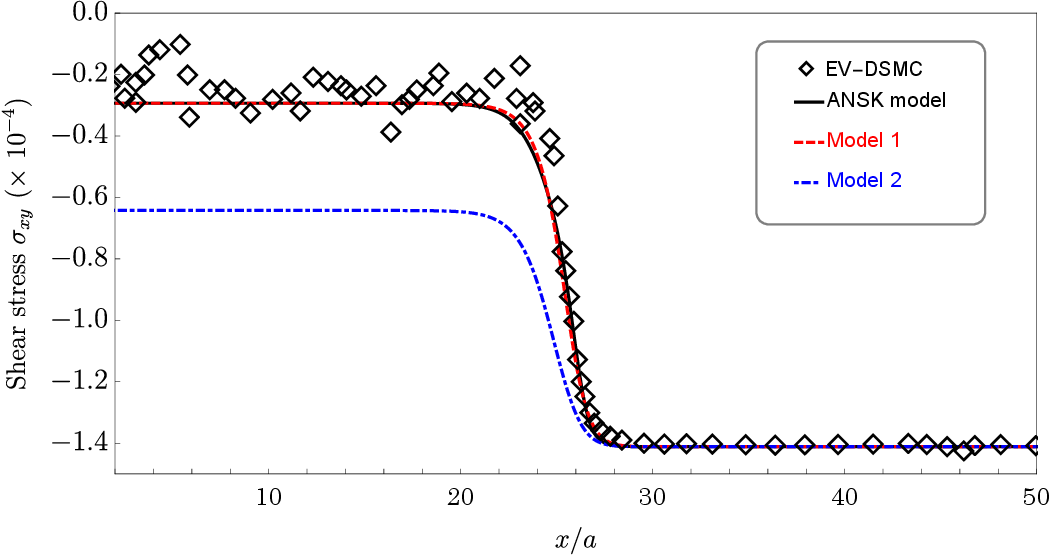}
\caption{\label{Couetteflowplots}\textcolor{black}{Two-phase Couette flow: comparison of EV-DSMC (black diamonds), ANSK model (black solid line), \textbf{Model 1} (red dashed line), and \textbf{Model 2} (blue dot-dashed line). Velocity profiles for \(\theta = 0.55, 0.60, 0.65\) (top row and bottom-left); shear stress at \(\theta = 0.60\) (bottom-right).}}
\end{figure}

Figure~\ref{Couetteflowplots} presents the velocity profiles \(v_y(x)\) for dimensionless temperatures \(\theta = 0.55\), \(0.60\), and \(0.65\), along with the shear stress \(\sigma_{xy}(x)\) at \(\theta = 0.65\), compared against EV-DSMC data (black diamonds). \textcolor{black}{The ANSK model (black solid line) shows excellent agreement with DSMC results, reflecting improved interface resistance from the temperature-dependent fitting of \(\alpha\) in the modified NSK framework.} In contrast, \textbf{Model 1} (\(\alpha = 0\)), which corresponds to the conventional NSK formulation, fails to resolve the sharp interfacial gradients observed at temperatures far from the critical point (red dashed line). Its performance improves as \(\theta\) approaches the critical point, where the interface becomes inherently smoother and thicker. \textcolor{black}{\textbf{Model 2} which omits the fourth order correction term in $a$ ---\(\bm{\Phi}^{(4)}\), results in notable discrepancies in both velocity and stress fields (blue dot-dashed line), underscoring the importance of this term.}

The shear stress is largely independent of \(\alpha\), depending instead on the higher-order correction. However, the inclusion of \(\alpha\) in the transport coefficients is essential for capturing sharp interface transitions at subcritical temperatures. \textcolor{black}{To determine a suitable temperature-dependent form of \(\alpha\), Couette flow simulations are conducted over a range of \(\theta\). A fitting procedure is then employed to match these results with solutions of the EV equation, following the approach outlined by \citet{Struchtrup_Frezzotti_2022}. This yields the following empirical relation for $\alpha$ as a function of temperature:
\begin{equation}
\alpha(\theta) = 9.5\theta - 4.25, \quad 0.45 < \theta < 0.7546.
\label{alphathetarelation}
\end{equation}
This fitted relation is then used in all subsequent analysis.}
\vspace{-0.3cm}
\subsection{\label{numericalschemeHT} \textcolor{black}{1D Steady-State Heat Transfer Without Phase Change}}
\textcolor{black}{Next, we consider steady-state, one-dimensional heat transfer across a liquid--vapor interface without phase change (\(J=0\)), where \(J\) denotes the total mass flux. The domain \(x \in [0, L]\) is discretized with a uniform grid spacing, with the interface located at \(x = L_0\). At the liquid boundary (\(x=0\)), a constant mass flux \(J\) and the temperature \(\theta_l\) are imposed, while at the vapor boundary (\(x=L\)), either the vapor temperature \(\theta_v\) is prescribed or, under adiabatic conditions, the temperature gradient---and thus the heat flux \(q_V\)---is set to zero, following \citet{struchtrup2024heat}.} The non-zero variables are
\{$\rho(x),q_{x}(x)={q}(x)$\}.

\textbf{Governing Equations:} Conservation laws for this case now reduce to 
\begin{align}
\label{oneheattransferwevep}
J = \rho v = 0, \quad
 \frac{d}{dx} \left[\rho \theta + \frac{2\pi}{3}  \rho^{2} \theta Y + \tau_{xx}^K\right] = 0, \quad -\lambda \frac{d\theta}{dx} = q_c = \text{constant},
\end{align}
where \(\tau_{xx}^{K}\) is the one-dimensional Korteweg stress component capturing capillarity effects. We solve Eqs.~\eqref{oneheattransferwevep} using a fourth-order finite difference scheme on a uniform grid with \(2n\) points, \textcolor{black}{incorporating the boundary conditions and an appropriate initial guess for both density and temperature profiles as outlined by \citet{struchtrup2024heat}. The resulting \(2n+1\) non-linear equations are then solved using the \texttt{fsolve} command in MATLAB\textsuperscript{\textregistered}.}

% \begin{align*}
% &\text{boundary \textsc{MATLAB}condition:} ~~~\rho(0) = \rho_l(\theta_{l}),\quad \rho(L) = \rho_v(\theta_{v}),\quad \frac{d\rho}{dx} = 0 \text{ at } x \neq L_0, \\
% &\text{fixed mass and pressure:} ~~~\int_0^L \rho \, dx = m_0 = \text{constant},\\
% &\text{initial guess (ig) for density profile:} ~~~\rho_{ig}(x) = \left(1 - \tanh(x - L_0)\right) \frac{\rho_l - \rho_v}{2} + \rho_v,\\
% &\text{initial guess for temperature profile:} ~~~\text{solutions of NSK equations from the bulk phase.}
% \end{align*}
\subsubsection{Results and Discussion}
\begin{figure}
    \centering
    \includegraphics[width = 0.50\textwidth]{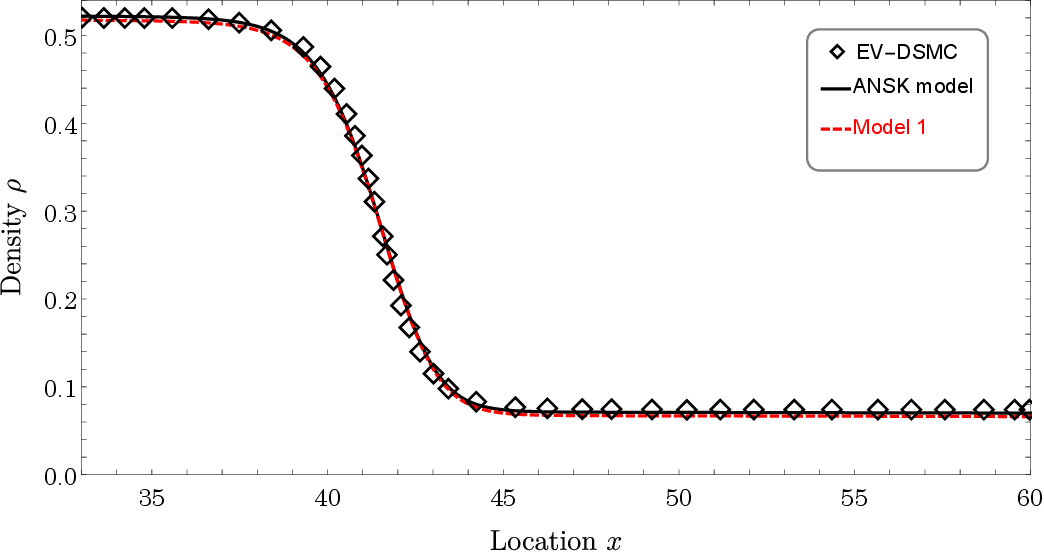}\hfill
     \includegraphics[width = 0.50\textwidth]{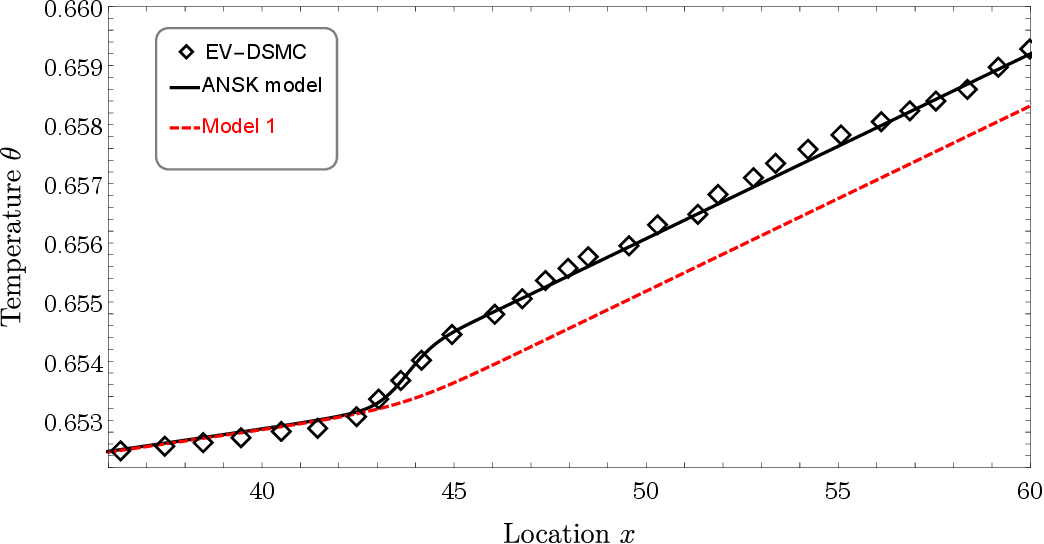}
     \includegraphics[width = 0.50\textwidth]{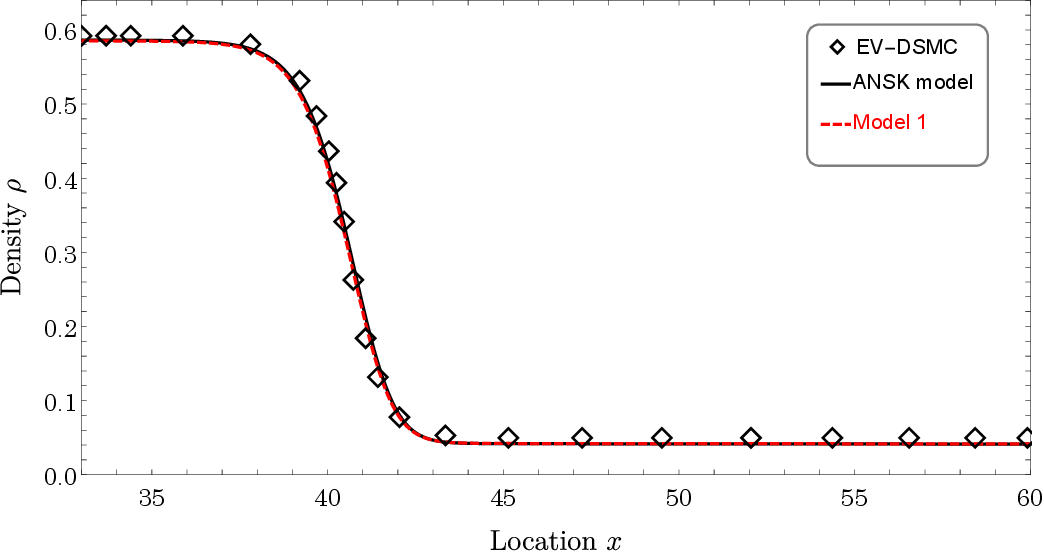}\hfill
     \includegraphics[width = 0.50\textwidth]{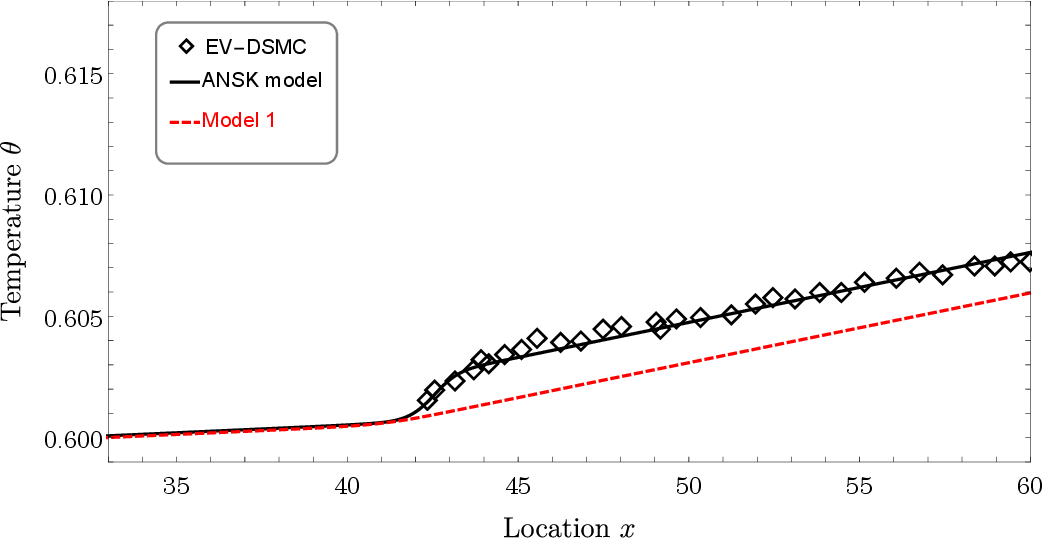}
     \includegraphics[width = 0.50\textwidth]{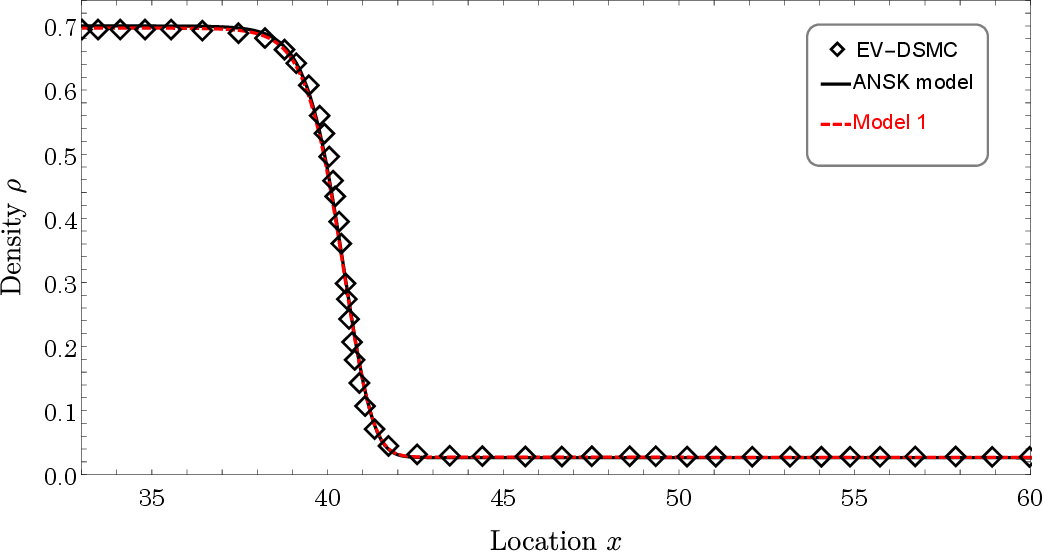}\hfill
     \includegraphics[width = 0.50\textwidth]{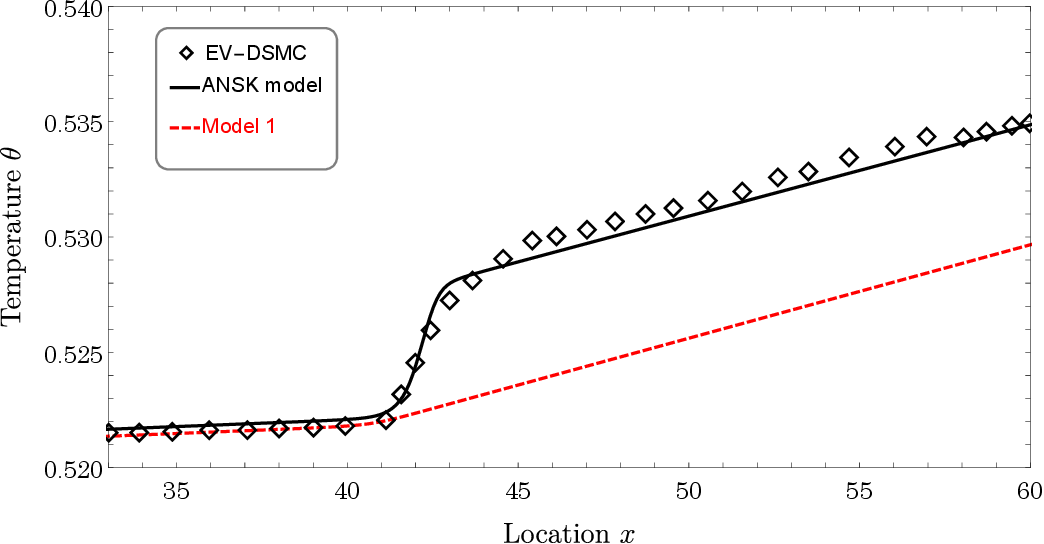}
\caption{\label{Heat transfer PlotsWevap}\textcolor{black}{Comparison of density and temperature from the ANSK model (black solid line), EV-DSMC (black diamonds), and \textbf{Model 1} (red dashed line) for three non-evaporative heat transfer cases. Rows: (i) $L=200$, $\theta_l=0.65$, $\theta_v=0.70$, $\alpha = 2.4$, $q_{c} = -1.90 \times 10^{-4}$; (ii) $L=120$, $\theta_l=0.60$, $\theta_v=0.64$, $\alpha = 1.83$, $q_{c} =- 1.55 \times 10^{-4}$; (iii) $L=220$, $\theta_l=0.52$, $\theta_v=0.60$, $\alpha = 0.69$, $q_{c} =- 1.99 \times 10^{-4}$; with $L_0=40$, $J=0$.}}
\end{figure}

Figure~\ref{Heat transfer PlotsWevap} shows density and temperature profiles for different boundary temperatures. The ANSK model (black solid line) shows excellent agreement with EV-DSMC data (black diamonds) \citep{struchtrup2024heat}, validating the role of the temperature-dependent capillarity coefficient \(\chi_3(\theta)\) \citep{bhattacharjee2024temperature} and gradient-dependent thermal conductivity \(\lambda(\rho, \nabla\rho)\) \citep{bedeaux2003nonequilibrium,johannessen2004nonequilibrium}. \textbf{Model 1} ($\alpha = 0$), which is basically the classical NSK equations, agrees well for the density profiles but fails to capture interfacial behavior and predicts flat interface for the temperature plots (red dashed line).
% Parameter values:
% \begin{table}
% \centering
% \caption{Boundary temperatures and corresponding model parameters.}
% \begin{tabular}{cccc}
% \toprule
% $\theta_l$ & $\theta_v$ & $\alpha$ & $q_c$ \\
% \midrule
% 0.65 & 0.70 & 2.40 & \(1.90 \times 10^{-4}\) \\
% 0.60 & 0.64 & 1.83 & \(1.55 \times 10^{-4}\) \\
% 0.52 & 0.60 & 0.69 & \(1.99 \times 10^{-4}\) \\
% \bottomrule
% \end{tabular}
% \end{table} 
\vspace{-0.4cm}
\subsection{{One dimensional steady state with forced evaporation}}
Evaporation strongly influences interfacial heat and mass transport. Following \citep{struchtrup2024heat}, we consider zero heat flux in the vapor bulk ($q_{\text{V}}$). The non-zero fields for this 1D problem with forced evaporation are
\{$\rho(x),\, v_{y}(x)=v_{y},\, \sigma_{xy}(x)=\sigma(x),\, {q_{x}}(x)={q(x)}$\}.

\textbf{Governing Equations:} Conservation laws for the case with forced evaporation reduce to 
\begin{equation}
\label{onedheat transferfevap}
\rho v = J = \text{constant},\quad
\frac{d}{dx} \left(\rho v_{y}^{2} + {\Pi_{x}}\right) = 0,\quad
\frac{d}{dx}\left[ \left( \frac{3}{2} \rho \theta + \frac{1}{2} \rho v_{y}^2 + \rho \varepsilon_{K,x} \right) v_{y} + \mathcal{J}_{x}\right] = 0,
\end{equation}
Here \(\Pi_x\), \(\varepsilon_{K,x}\) and \(\mathcal{J}_x\) denote the \(x\)-components, in the one-dimensional setting, of the total momentum flux tensor, nonlocal potential energy, and energy flux vector as defined above.

Equations (\ref{onedheat transferfevap}) are solved using the numerical method outlined in (\ref{numericalschemeHT}), with boundary conditions at both the left and right boundaries specified based on the EV-DSMC data. These boundary conditions are then used to solve the NSK equations in the bulk, with the resulting solutions serving as initial guess profiles for the equations in (\ref{onedheat transferfevap}).
\subsubsection{Results and Discussion}
\begin{figure}
    \centering
    \includegraphics[width = 0.50\textwidth]{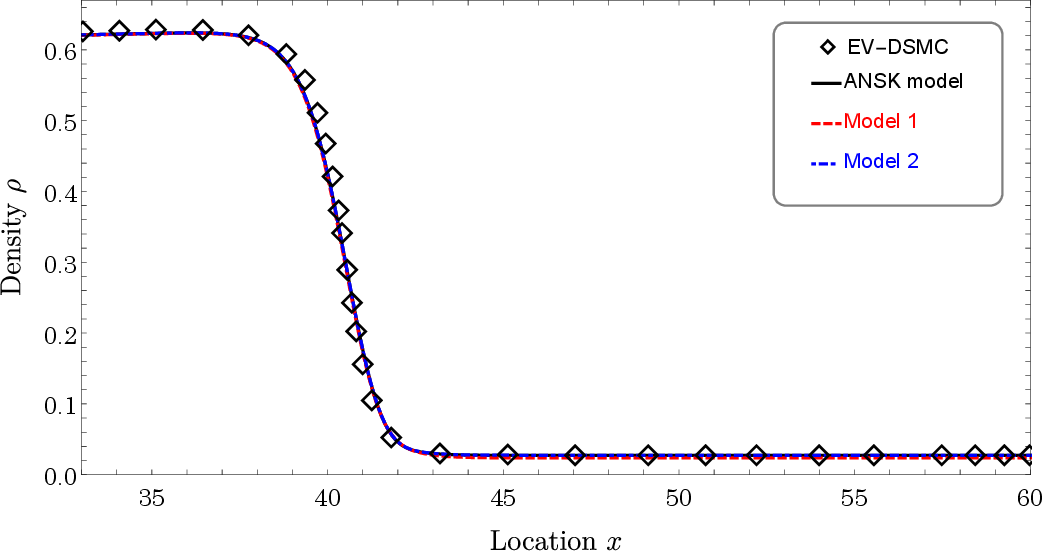}\hfill
      \includegraphics[width = 0.50\textwidth]{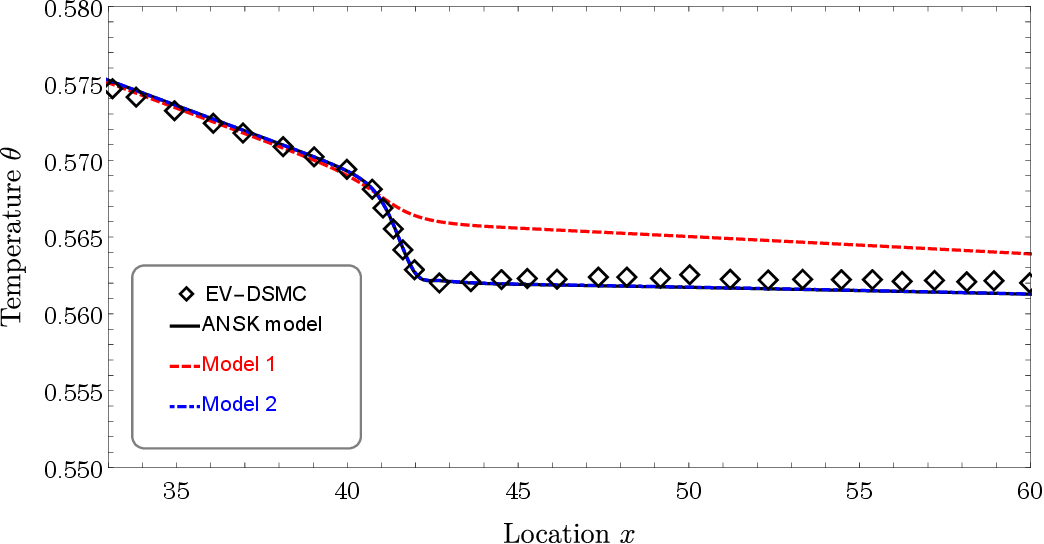}
     \includegraphics[width = 0.50\textwidth]{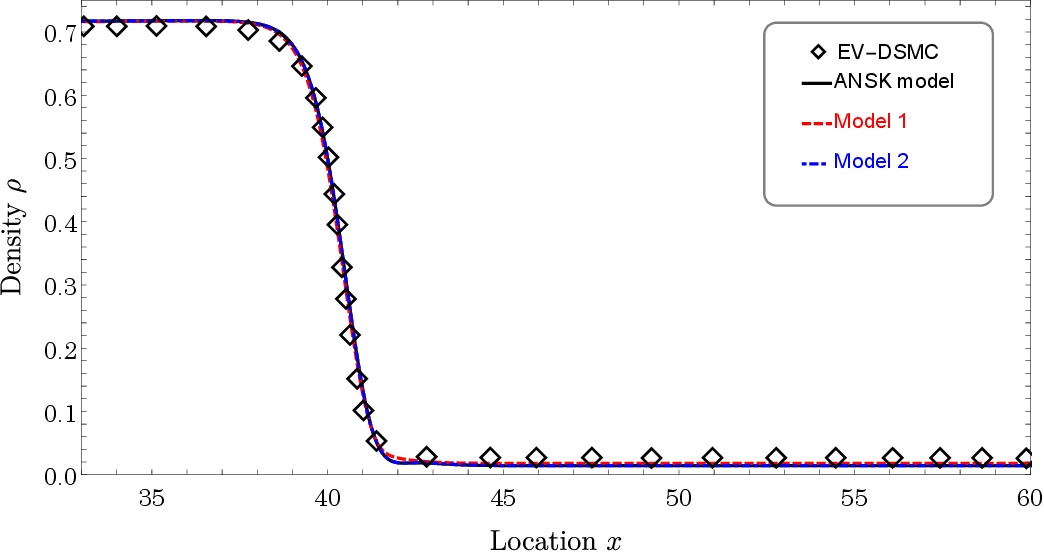}\hfill
     \includegraphics[width = 0.50\textwidth]{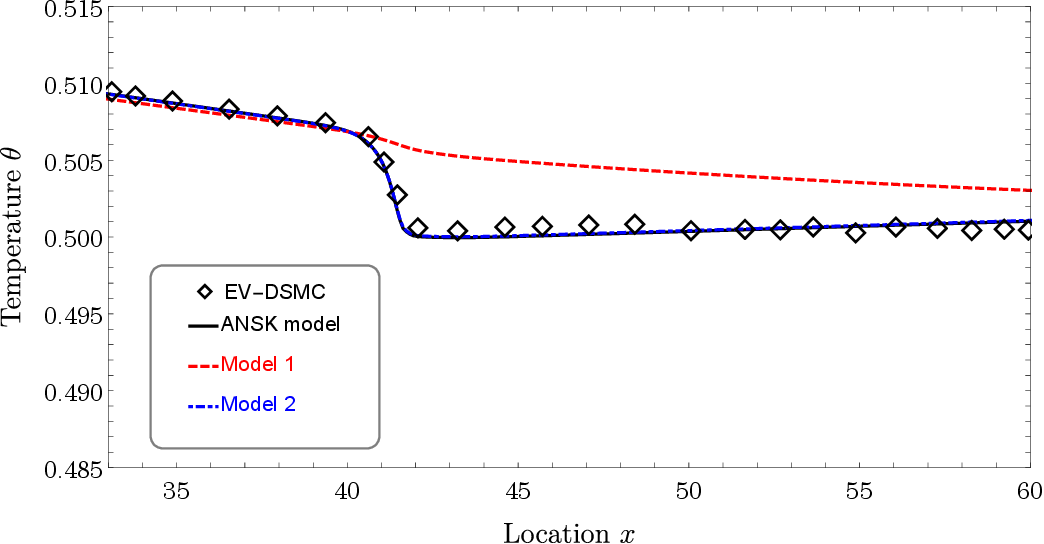}
\caption{\label{Heat transfer Plotsevap}Density and temperature profiles with forced evaporation, comparing the ANSK model, EV-DSMC, \textbf{Model 1}, and \textbf{Model 2}. Top: $L = 200$, $\theta_l = 0.60$, $\theta_v = 0.56$, $J = 0.000745$, $\alpha = 1.26$. Bottom: $L = 220$, $\theta_l = 0.52$, $\theta_v = 0.48$, $J = 0.00035$, $\alpha = 0.31$. Interface at $L_0 = 60$.}
\end{figure}

Figure~\ref{Heat transfer Plotsevap} shows density (left panel) and temperature (right panel) profiles for two evaporation cases, with $L_0 = 60$, and $q_{\text{V}} = 0$. 
Top row: $\theta_l = 0.60$, $\theta_v = 0.56$, $J = 0.000745$, $\alpha = 1.26$;  
Bottom row: $\theta_l = 0.52$, $\theta_v = 0.48$, $J = 0.00035$, $\alpha = 0.31$.
The ANSK model (black solid lines) agrees closely with EV-DSMC results (black diamonds) \citep{struchtrup2024heat}, for both the density and temperature profile. \textbf{Model 2} results coincides with the ANSK model showing negligible role of stress gradient in heat transfer problems.
\vspace{-0.5cm}
\section{\label{sec:level5}Conclusion}
\textcolor{black}{This work presents a thermodynamically consistent extension of the classical Navier-Stokes-Korteweg (NSK) model, systematically derived from the EV kinetic equation to accurately capture interfacial transport processes at the nanoscale. While the classical Navier–Stokes–Korteweg (NSK) equations assume local equilibrium and constant transport properties—assumptions that break down in regions with steep gradients and strong non-equilibrium effects—the proposed model incorporates microscopic physics through a continuum framework. This makes it suitable for simulating dense fluids and multiphase systems where classical formulations fall short.}

\textcolor{black}{Validation against direct simulation Monte Carlo (DSMC) \textcolor{black}{for the Enskog-Vlasov equation,} for three benchmark problems---two-phase Couette flow, non-evaporative heat conduction, and planar evaporation---demonstrates that the ANSK model achieves close agreement with kinetic theory predictions. In particular, the inclusion of temperature-dependent Korteweg coefficient, density gradient-dependent transport coefficient with temperature-dependent parameter $\alpha$, in the ANSK model significantly improves its accuracy in capturing interfacial dynamics, without compromising computational efficiency. Moreover, the model adheres to the second law of thermodynamics, as confirmed through a detailed entropy production analysis.}

\textcolor{black}{Overall, this study bridges the gap between computationally expensive kinetic theory models and continuum-based approaches, offering a robust framework for applications in nanoscale thermal management, phase-change heat transfer, and microfluidic systems.}

\textcolor{black}{
Future directions include the exploration of thermodynamically consistent boundary conditions, the development of more robust numerical schemes, and applications to capillary formation in nanotubes. Most importantly, since the ANSK equations cannot capture Knudsen layers and other non-equilibrium effects---such as parallel heat flux in Couette flow that deviates from Fourier’s law---more accurate models, as presented in the literature \cite{Struchtrup_Frezzotti_2022}, will also be explored.}
% \backsection[Declaration of Competing Interest]{The authors declare that they have no known competing financial interests or personal relationships that could have appeared to influence the work reported in this paper.}
% \backsection[Acknowledgements]{HS acknowledges support from NSERC (Discovery Grant RGPIN-2022-03188). ASR acknowledges support from SERB, India (MATRICS MTR/2021/000417).}

\bibliographystyle{plainnat}
\bibliography{jfm}

\end{document}